\documentstyle[12pt,epsfig]{article}
\author{H. Rafii-Tabar$^1$
\thanks{Corresponding Author: e-mail: rafii-tabar@nano.ipm.ac.ir,
Tel:(+98)212835058,Fax:(+98)
212280415.}\,\,, H. R.
Sepangi$^{1,2}$\\
{\small $^1$ Computational Physical
Sciences Research Laboratory, Department of Nano-Science,}\\
{\small Institute for Studies
in Theoretical Physics and
Mathematics (IPM), P.O. Box 19395-
5531,Tehran, Iran.}\\ {\small $^2$
Department of Physics, Shahid Beheshti
University, Evin, Tehran 19839,
Iran.}}
\title{\bf Numerical simulation of the stochastic dynamics of inclusions in biomembranes in presence of surface tension}
\begin{document}
\maketitle
\vspace{11cm}
\pagebreak
\begin{abstract}
The stochastic dynamics of inclusions in a randomly fluctuating biomembrane is simulated. These inclusions can represent the embedded proteins and the external particles arriving at a cell membrane. The energetics of the biomembrane is modelled via the Canham-Helfrich Hamiltonian. The contributions of both the bending elastic-curvature energy and the surface tension of the biomembrane are taken into account. The biomembrane is treated as a two-dimensional sheet whose height variations from a reference frame is treated as a stochastic Wiener process. The lateral diffusion parameter associated with this Wiener process coupled with the longitudinal diffusion parameter obtained from the standard Einsteinian diffusion theory completely determine the stochastic motion of the inclusions. It is shown that the presence of surface tension significantly affects the overall dynamics of the inclusions, particularly the rate of capture of the external inclusions, such as drug particles, at the site of the embedded inclusions, such as the embedded proteins.
\vspace{10mm }
\end{abstract}
\pagebreak
\section{Introduction}
Amphiphilic molecules, such as lipids and proteins, are composed of hydrophilic heads and hydrophobic hydrocarbon-based chains. They can self-assemble themselves into a variety of exotic structures when placed in an aqueous environment \cite{Hunt}. One such structure is the biological membrane composed of a bilayer of phospholipid molecules. Biomembranes have very interesting spatial dimensions.
Their thicknesses can vary up to few
nano-meters, but their linear sizes can
take up to tens of micro-meters. They
can, therefore, be regarded as highly
flexible {\it fluid-like} two-dimensional (2D)
sheets embedded in three-dimensional
embedding space.

Biomembranes form
large encapsulating bags, called {\it
vesicles}, because open sheet-like
configurations would involve a large
energy along the hydrophobic
edges \cite{Seifert1}. They play a very
significant role in many of the life's
processes, since not only they act as
barriers to maintain the structural
integrity of cells, but also provide
the functional environment for many of
the embedded proteins that penetrate
the biomembrane thicknesses and act as ion
transport channels to the interior of the
cells \cite{Nico}.

Biomembranes can be studied from
three perspectives \cite{Seifert1}, namely
(i) their molecular architecture, focusing
on their material properties, (ii) as
statistical-mechanical systems displaying
a very rich and exotic variety of
configurations as 2D surfaces, and (iii)
their functioning in biological systems.

In our previous work \cite{Hashem}, as
well as in this paper, we have been
interested in biomembranes from the second
perspective (ii), and in particular we
have been concerned with computational modelling of the
diffusion dynamics of objects
(inclusions) existing within and on the surface of stochastically
fluctuating biomembranes. The stochastic
fluctuations of the biomembranes can originate from thermal
fluctuations. They promote shape
fluctuations and shape transformations
in biomembranes. One such shape
transformation, observed under video
microscopy, is the so-called {\it budding}
transition in which a change of
temperature from $T=27.2$ C to $T=41
$ C resulted in a change of shape of a
spherical vesicle into a prolate ellipsoid,
and the eventual fission of the
biomembrane \cite{Seifert1}.

Two types of inclusion can be considered in
a biomembrane. These are the {\it internal} inclusions, or the biomembrane's own inclusions, which we refer to as the
M-type, and  the {\it external} inclusions, which we refer to as the S-type. The S-type inclusions, such as drug particles, arrive at the biomembrane from outside. The M-type inclusions are the {\it embedded} objects, such as proteins, or ion channels, that
are an integral part of a biomembrane. Although embedded in the amphiphilic environment, these inclusions are {\it mobile} and can freely diffuse across the biomembrane.

The M-type inclusions penetrate the thickness of the biomembrane, and hence disturb its geometry locally by forcing the amphiphilic bilayer to adjust its thickness locally so as to match its hydrophobic regions with those of the inclusions \cite{Dan1, Dan2}, as shown in Fig.1. This change in the local geometry generates
perturbations in the fluid-like biomembrane
structure and these perturbations, in turn, lead to both short- and long-range {\it
biomembrane-induced} forces between the inclusions. Such geometro-dynammic forces affect the behaviour of inclusions in a biomembrane {\it in addition} to any other forces, such as direct molecular forces, that may operate between these
inclusions, or between them and the embedding biomembrane. If these induced
perturbations are of long wavelength, then they generate long-range forces,
otherwise, they lead to short-range forces that act in the immediate vicinity of the
inclusions.

The deformation in local geometry is one
of the {\it three} deformation modes of a biomembrane.
Since a biomembrane is surrounded by solutions,
such as water, the {\it surface tension}
existing at the biomembrane-water interface
changes the overall surface area of the
biomembrane. This change constitutes the
second mode of deformation. The third
mode is associated with the {\it bending}
property of a biomembrane, which is
associated with its elastic-curvature
property. It is this property that
distinguishes a fluid-like biomembrane
sheet from a simple fluid sheet in which
surface tension alone dominates the
surface dynamics.
\cite{Seifert1,Mansfield}.

In the studies concerned with the
inclusion-induced local deformations, two
approaches have been adopted to model
these deformations. In one
model \cite{Bloom,Abney,Marcelja,
Owicki, Fattal} the biomembrane energy is
taken to be the sum of the energy due
to molecular compression/expansion
brought about by the change in local
geometry and the energy due to the change in the
overall surface area, while in another
model \cite{Dan1,Huang} the
contribution
of the bending energy is also included.
The first model shows that the inclusion-
induced deformations force a change of
thickness in local geometry which decays
{\it exponentially} from the
\mbox{inclusion-imposed} value to the
equilibrium
thickness of the biomembrane. This is
shown schematically in Fig.1
for two rod-like inclusions. The second
model shows that the contribution of
bending energy could significantly affect
the deformation profile at the inclusion
boundary, as well as the biomembrane-induced interactions.

The S-type inclusions arrive
from the outside of the biomembrane and
reside on its {\it surface}. Fig.2
represents a schematic representation of
two such inclusions lying on the surface
of a biomembrane sheet.

In our previous work \cite{Hashem},
we performed a computational simulation
of the diffusion dynamics of both the external
and internal inclusions in a
stochastically fluctuating biomembrane in
which the biomembrane energy was solely
described in terms of the bending \mbox{
elastic-curvature} energy. Furthermore, in that work we considered only the contribution of the lateral component of the diffusion coefficient of the inclusions, brought about by the lateral fluctuations in the height of the biomembrane, to the motion of the inclusions.  In the present paper, we have extended our previous modelling on two levels. Firstly, have also included the contribution of
biomembrane's surface tension to the
stochastic dynamics of inclusions, and secondly, the contribution of the longitudinal component of the diffusion coefficient has also been included. The combination of these two components now forms the diffusion coefficient governing the motion of the inclusions. Various forms of this combination has been tested. It is seen that the dynamics of inclusions is significantly affected by the presence of surface tension.
\section{Energetics of the biomembrane in presence of surface tension}
Biomembranes are characterized by their bending rigidity
\cite{Stat, Safran1, Gompper} and they
strongly undulate under thermal
fluctuations. Consequently, in studying
biomembranes, surface tension effects are
usually neglected and only the
bending rigidity is considered as the
source controlling their shape and
fluctuations. However,
surface tension can
contribute significantly to the energy of
a biomembrane if geometrical constraints
\cite{Seifert2}  or external perturbations,
due to the attached proteins in biological
systems, are introduced
\cite{Alberts}. For instance, it has been
shown that \cite{Sens} the area
dependence of surface tension can
lead to a stable hole of finite size in the
biomembrane. Furthermore, interesting
phenomena have been
observed when biomembranes are perturbed
by the action of microscopes
\cite{Evans}, or in recent years by laser
tweezers \cite{Bar}. The typical effect of
such tweezers on an initially flaccid
biomembrane has been reported \cite{Bar}
where the application of the laser
tweezers on a vesicle over few minutes
caused the shape fluctuations of the
biomembrane to disappear. With this
disappearance, the biomembrane became
quite {\it taut}, indicating the presence
of a non-negligible surface tension,
$\sigma\ge 10^{-3}$erg/cm$^2$. This
tension caused the expulsion of a vesicle
from inside of another vesicle. Such an
spontaneous expulsion is quite
surprising \cite{Sens}, since it involves
the opening of a large hole, of energy
much larger than the thermal energy.
In comparison, using mechanical
manipulation, rather than tweezers, of
very large vesicles, tensions as low as,
$\sigma\ge 10^{-6}$erg/cm$^2$
\cite{Evans} have also been recorded. It
is, therefore, quite reasonable to assume that
surface tension in the above range plays
a significant role in the dynamic
behaviour of a biomembrane.
These experiments, therefore, show that
a realistic study of the energetics of biomembranes
should include not only the bending
rigidity, but also the effect of the
surface tension.

The free elastic energy
of a symmetric, nearly flat, biomembrane sheet
is described by the Canham-Helfrich
Hamiltonian
\cite{Can,Hel}
\begin{eqnarray}
{\cal H}=\int
d^2\sigma\sqrt{g}\left\{\sigma_0+2
\kappa
H^2+\bar{\kappa} K\right\},
\label{eq1}
\end{eqnarray}
where
\begin{eqnarray}
K&=&\mbox{det}(K_{ij})=\frac{1}{R_1R
_2},\nonumber\\
H&=&\frac{1}{2}\mbox{Tr}(K_{ij})=
\frac{1}{2}\left(\frac{1}{R_1}+
\frac{1}{R_2}\right), \nonumber
\end{eqnarray}
are the Gaussian and mean curvatures of
the sheet respectively,
$R_1$ and $R_2$ are the principal radii
of the sheet, $\sigma_0$
is the surface tension, $\kappa$ is the
bending rigidity,
$\bar{\kappa}$ is the Gaussian rigidity,
$g$ is the determinant of
the metric tensor and
$\sigma=(\sigma_1,\sigma_2)$ is the
local coordinate defined on the sheet.
The last term on the RHS of (\ref{eq1}) is, by
Gauss-Bonnet
theorem, a topological invariant and
does not affect the
dynamics of the biomembrane if there were no
changes in its topology. We will, therefore, ignore this term as we
assume that no such changes occur. We
thus have
\begin{eqnarray}
{\cal H}=\int
d^2\sigma\sqrt{g}\left\{\sigma_0+2
\kappa
H^2\right\}. \label{eq2}
\end{eqnarray}

To facilitate the study of a nearly flat
biomembrane, whose free
energy is given by (\ref{eq2}), it is
convenient to consider
it to be parallel to the $(x_1-x_2)$ plane,
regarded as the reference plane.
The position of a point on the biomembrane
can then be described by a
single-valued {\it height} function,
$h$, representing the position of
a point on a fluctuating, nearly flat,
sheet relative to the reference plane. In
this, the so-called {\it Monge
representation} \cite{Seifert1},
equation (\ref{eq2}) is written as
\begin{eqnarray}
{\cal H}=\frac{1}{2}\int
d^2x\left\{\kappa(\nabla^2
h)^2+\sigma_0(\nabla h)^2\right\}.
\label{eq3}
\end{eqnarray}
In this form, the Hamiltonian is
expressed
solely in terms of the
height function, $h$, and its derivatives.
We remark that in our previous work \cite{Hashem} the
contribution of the second term on the
RHS of (\ref{eq3}) was not included as we did not consider the contribution of the surface tension $\sigma_0$.
The Hamiltonian in (\ref{eq3}) describes the energetics of the biomembrane from which we obtain the stochastic {\it lateral} motion of both the M-type and the S-type inclusions.
\subsection{Computation of the lateral diffusion coefficient of inclusions}
Let us first consider the lateral stochastic motion of the inclusions, i.e. that component of the motion which is associated with the changes in $h(x_1,x_2)$. One way to introduce stochastic behaviour into the dynamics of the biomembrane is to treat $h$, obtained from (\ref{eq3}), as a stochastically fluctuating Wiener process. Evidently, this stochastic behaviour is communicated to the inclusions residing inside, as well as on the surface, of the biomembrane.

To derive the lateral diffusion coefficient, $D_{\bot}$, associated with the random changes in the height function, we first need to obtain the the height-height correlation function
of the biomembrane in real space. This is obtained from (\ref{eq3}) by first going over to the
Fourier representation
of $h$, i.e.
\begin{eqnarray}
h({\bf x})=\int\frac{d^2 q}{(2\pi)^2}h(\bf
q)e^{i{\bf q}\cdot{\bf
x}}\,,\label{eq4}
\end{eqnarray}
where ${\bf x}=(x_1,x_2)$ is the vector position of the point on the biomembrane.
Consequently, (\ref{eq3}) now reads
\begin{eqnarray}
{\cal H}=\frac{1}{2}\int\frac{d^2
q}{(2\pi)^2}\left\{\kappa q^4+
\sigma_0  q^2\right\}h({\bf
q})h^\ast({\bf q})\,, \label{eq5}
\end{eqnarray}
where $*$ denotes the complex conjugate.
The {\it static} height-height correlation
function can be
calculated, with the result
\begin{eqnarray}
\langle h({\bf q};0)h^\ast({\bf
q^\prime};0)\rangle=\left(\frac{k_{\tiny\mbox{B}} T }
{\kappa q^4+\sigma_0
q^2}\right)(2\pi)^2\delta({\bf q}-{\bf
q}^\prime), \label{eq6}
\end{eqnarray}
where the averaging is done with respect
to the Boltzmann weight
factor $\mbox{exp}(-{\cal H}/k_{\tiny\mbox{B}} T)$,
with $k_{\tiny\mbox{B}}$ being the
Boltzmann constant. The corresponding
{\it dynamic} correlation function
can be written as
\begin{eqnarray}
h({\bf q};t)=h({\bf
q};0)e^{-\gamma(q)t},\nonumber
\end{eqnarray}
leading to
\begin{eqnarray}
\langle h({\bf q};t)h^\ast({\bf
q^\prime};0)\rangle=
e^{-\gamma(q)t} \left(\frac{k_B T }
{\kappa q^4+\sigma_0
q^2}\right)(2\pi)^2\delta({\bf q}-{\bf
q}^\prime)
 , \label{eq7}
\end{eqnarray}
where the damping factor, $\gamma(q)$,
reflecting the long-range
character of the hydrodynamic damping,
is given by \cite{Seifert1}
\begin{eqnarray}
\gamma(q)=\frac{1}{4\eta}\left(\kappa
q^3+\sigma_0 q\right).
\label{eq8}
\end{eqnarray}
Here, $\eta$ denotes the coefficient of
viscosity of the fluid surrounding the
biomembrane. We note that in the absence
of $\sigma_0$, (\ref{eq8})
reduces to the damping factor given in
equation (13) of our previous work
\cite{Hashem}.
In real space, the Fourier transform of (\ref{eq7}) is given by
\begin{eqnarray}
\langle h({\bf x};t)h({\bf x};0)\rangle=\int\int\frac{d^2 q}{(2\pi)^2}\frac{d^2 q^{\prime}}{(2\pi)^2}\langle h({\bf q};t)h^*({\bf q^{\prime}};0)\rangle e^{-{\bf x}.({\bf q}-{\bf q^{\prime})}}\,,\label{eq8a}
\end{eqnarray}
or substituting from (\ref{eq7}) and (\ref{eq8})
\begin{equation}
\langle h({\bf x};t)h({\bf x};0)\rangle=\frac{1}{2\pi\beta}\int_{\frac{1}{L}}^{\frac{1}{a}}dq \frac{e^{-\frac{1}{4\eta}(\kappa q^3+\sigma_0q)t}}{\kappa q^3+\sigma_0 q}\,,
\label{eq9}
\end{equation}
where $L$ is the linear size of the
biomembrane and $a$ is a
molecular cut-off, of the order of
nanometers, and $\beta=\frac{1}{k_{\tiny\mbox{B}}T}$.
The Fourier transform expression in (\ref{eq9}) is quite complicated to perform and  it was computed in the Mathematica software \cite{Mathematica} by expanding the integrand and computing it term by term. The  final analytical result, consisting of many terms, need not be quoted here and can be obtained from the authors upon request.
In our previous work \cite{Hashem}, the Fourier transform to real space was performed under simplifying assumptions regarding the transform of the damping factor term $\gamma$. In this paper, however, no assumptions were applied and the transform was fully computed.

Now, treating $h$ as a stochastic
Wiener process implies that
the mean and variance are given by
\begin{eqnarray}
\langle h(x_1,x_2;t)\rangle&=&0,
\label{eq10}\\ \langle
h(x_1,x_2;t)h(x_1,x_2;0)\rangle&=&
=2D_{\bot}t, \label{eq11}
\end{eqnarray}
in which $D_{\bot}$,  the diffusion constant
associated with the random fluctuations in the height
function at the local position $(x_1,x_2)$,
represents a measure of these fluctuations.
Such random fluctuations can cause a
roughening of the biomembrane surface as has been
observed in NMR experiments \cite{Chiu}.

Consequently, since the biomembrane's stochastic behaviour is
transmitted to the inclusions, we can make the assumption that the centre of mass of
an inclusion coinciding
with the point $(x_1,x_2)$ on the
biomembrane would also experience the
same lateral fluctuations, whose measure is
the same diffusion coefficient, $D_{\bot}$.

Therefore, the expression for $D_{\bot}$ for the inclusions
in presence of surface tension
can be obtained by comparing
(\ref{eq9}) and (\ref{eq11}), leading to
\begin{equation}
D_{\bot}=\frac{1}{2t}\left[\frac{1}{2\pi\beta}\int_{\frac{1}{L}}^{\frac{1}{a}}dq \frac{e^{-\frac{1}{4\eta}(\kappa q^3+\sigma_0q)t}}{\kappa q^3+\sigma_0 q}\right]. \label{eq12}
\end{equation}
\subsection{Computation of the longitudinal diffusion coefficient of inclusions}
In addition to the lateral component of the diffusion coefficient, computed above, we also require the longitudinal component, $D_{\|}$, for both the S-type and the M-type inclusions. This coefficient is calculated via Einstein's general model of Brownian dynamics \cite{Haile} according to which
\begin{equation}
D_{\|}=\frac{1}{2Nst}\langle\sum_{i=1}^N\mid r_i(t)-r_i(0)\mid^2\rangle\,,\label{eq12a}
\end{equation}
where $t$ is the {\it delay} (correlation) time, $\langle\cdots \rangle$ refers to averaging over time, $s$ is the dimensionality of diffusion space (2 in the present case) and $N$ is the number of inclusions. In the actual computation of $D_{\|}$, the time average is replaced with a summation over time origins from which the delay time is measured, so that (\ref{eq12a}) is written as \cite{Haile}
\begin{equation}
D_{\|}=\frac{1}{2XNst}\sum_{k=1}^{X}\sum_{i=1}^{N}\left[\mid r_i(t_k+t)-r_i(t_k)\mid^2\right]\,,\label{eq12b}
\end{equation}
where $X$ is total number of available time origins.
\subsection{Computation of the stochastic trajectories}
In our present numerical simulations,
just as in the previous study
\cite{Hashem}, the equations of motion
of both the M-type and the S-type
inclusions are described via the
Ito stochastic calculus \cite{Ito} whose
stochastic differential equation is given
by
\begin{eqnarray}
d{\bf r}(t)={\bf A}[{\bf
r}(t),t]+D_t^{1/2}d{\bf W}(t). \label{eq13}
\end{eqnarray}
This equation describes the random
trajectories, ${\bf r}(t)$,
of the centres of mass of the inclusions in
terms of a dynamical
variable of the inclusions, ${\bf A}[{\bf
r}(t),t]$, which is
referred to as the drift velocity, and a
term, $d{\bf W}(t)$,
which is modelled by a Wiener process with
the mean and variance
given by
\begin{eqnarray}
\langle d{\bf W}(t)\rangle&=&0,\hspace{3mm}\nonumber\\
\langle d{\bf W}_i(t)d{\bf W}_j(t)\rangle&=&2\delta_{ij}dt.
\label{eq14}
\end{eqnarray}
In (\ref{eq13}), $D_t$ represents the total diffusion coefficient of the inclusions. In this paper, following the prescription given by \cite{Dhont} for the Brownian motion of rod-like objects, we represent this coefficient by  a linear combination of the lateral and longitudinal coefficients and, to test the various possibilities, consider the following different combinations
\begin{eqnarray}
D_t&=&\frac{1}{2} D_{\bot}+\frac{1}{2}D_{\|}\nonumber\\
&=&\frac{2}{3}D_{\bot}+\frac{1}{3}D_{\|}\nonumber\\
&=&\frac{1}{3}D_{\bot}+\frac{2}{3} D_{\|}.\label{eq14a}
\end{eqnarray}

In the vicinity of an M-type inclusion, care should be exercised  in using $D_t$. Near an M-type inclusion, localized exponentially-decaying deformation would occur in the membrane geometry \cite{Hashem, Dan1}, as depicted in Fig.1. This implies that near such an inclusion we can scale the diffusion coefficient by an exponentially decaying function, i.e. if $r_0$ is the radius of an M-type inclusion, then within a circular region $R+r_0$, the modified diffusion coefficient could be modelled as
\begin{equation}
D_{mt}=D_{t} e^{\frac{-r_0}{R}}\,,\label{eq14b}
\end{equation}
implying that when an S-type inclusion enters a region centered around an M-type inclusion, within the radius $R+r_0$, the diffusion coefficient goes over to $D_{mt}$ and, to a first approximation, this is how an M-type inclusion interacts with an S-type one.

The Ito equation (\ref{eq13}) predicts the increment
in position for a
meso-scale time interval, $dt$, as a
combination of  deterministic
and stochastic diffusive parts
represented by the terms ${\bf
A}[{\bf r}(t)]$ and $D_t^{1/2}d{\bf W}(t)$
respectively. This
equation resembles the `position'
Langevin equation describing the
Brownian motion of a particle
\cite{Allen}. The position Langevin
equation corresponds to the {\it
long-time} (diffusive time)
configurational dynamics of a stochastic
particle in which its
momentum coordinates are in thermal
equilibrium and hence can
be removed from the equations of
motion. Since we are interested
in diffusive time scales as well, we can
re-write (\ref{eq13})
as
\begin{eqnarray}
d{\bf r}(t)=\frac{D_t}{k_B T}{\bf
F}(t)+D_t^{1/2}d{\bf W}(t),
\label{eq15}
\end{eqnarray}
where ${\bf F}(t)$ is the instantaneous
systematic (Newtonian) force
experienced by the centre of mass of the
$i$th inclusion. This force can obtained
from the
inter-inclusion potentials, $V(R_{ij})$,
according to
\begin{eqnarray}
{\bf F}_i=-\sum_{j>i}\nabla_{{\bf
R}_i}V(R_{ij}). \label{eq16}
\end{eqnarray}
Pertinent potentials have been
constructed
employing a statistical mechanics
based on the
Canham-Helfrich Hamiltonian. For
the M-M interaction, this potential
is given by \cite{Goul1,Goul2}
\begin{eqnarray}
V^T_{\tiny\mbox{MM}}(R_{ij})=-k_B
T\frac{12A^2}{\pi^2R_{ij}^4},
\label{eq17}
\end{eqnarray}
describing the biomembrane-mediated
temperature-dependent long-range
interactions between a pair of disk shape
M-type inclusions that can
freely tilt with respect to each other. For
the S-S interaction, the
potential is given by \cite{Gole}
\begin{eqnarray}
V^T_{\tiny\mbox{SS}}(R_{ij},\theta_i,\theta_j)=-k_B
T\frac{L_i^2 L_j^2}{128
R^4_{ij}}\cos^2[2(\theta_i+\theta_j)],
\label{eq18}
\end{eqnarray}
where $A=\pi r_0^2$ is the area of an
M-type inclusion of radius
$r_0$, $R_{ij}$ is the distance between
the centres of mass of two
inclusions $i$ and $j$, $L_i$ and $L_j$
are the lengths of two
S-type inclusions making angles
$\theta_i$ and $\theta_j$
respectively with the line joining their
centres of mass (Fig.2) and $T$
is the biomembrane temperature. These
expressions are derived for the
rod-like inclusions that are assumed to
be much more rigid than
the ambient biomembrane so that these
inclusions cannot move
coherently with the biomembrane.

We performed numerical simulations of the space-time motion of inclusions
described by (\ref{eq15})
by adopting the following iterative
scheme \cite{Hashem1}
\begin{eqnarray}
X(t+dt)=X(t)+\frac{D_t}{k_B
T}F_X(t)dt+\sqrt{2D_tdt}\,
R^G_X,\nonumber\\
\label{eq19}\\
Y(t+dt)=Y(t)+\frac{D_t}{k_B
T}F_Y(t)dt+\sqrt{2D_tdt}\,
R^G_Y,\nonumber
\end{eqnarray}
where $R^G_X$ and $R^G_Y$ are
standard random Gaussian variables
chosen separately and independently for
each inclusion according
to the procedure given in \cite{Allen},
and $F_X$, $F_Y$ are the
$X$ and $Y$ components of the force
$\bf F$. For the S-type
inclusions, we treated the angles in
(\ref{eq18}) as
independent stochastic variables
described by
\begin{eqnarray}
\theta(d+dt)=\theta(t)+\frac{D_t}{k_B
TL^2}\tau(t)dt+\frac{1}{L}\sqrt{2D_tdt}
\,\theta^G, \label{eq20}
\end{eqnarray}
where $\tau$ is the torque experienced
by an S-type inclusion and
is given by
\begin{eqnarray}
\tau_i=-\sum_{j>i}\frac{\partial
V^T(R_{ij},\theta_i,\theta_j)}{\partial
\theta_i}, \label{eq21}
\end{eqnarray}
and $\theta^G$ is the angular
analogue of $R^G_X$ and $R^G_Y$.
\section{Results and discussion}
In our simulations, we used a square membrane patch together with the following data:
\begin{eqnarray}
L&=&40\,\mu\mbox{m}\nonumber\\
\kappa&=&10^{-19}\,
\mbox{J}\nonumber\\
\eta&=&10^{-3}\,\mbox{J\, s\,}
\mbox{m}^
{-3}\nonumber\\
\sigma_0&=&5\times 10^{-
3}\mbox{erg}/\mbox{cm}^2\,\, \mbox{to}
\,\,5\times 10^
{-
6}\mbox{erg}/\mbox{cm}^2\nonumber\\
T&=&300 \mbox{K}\nonumber\\
m&=&10^{-12}\mu\mbox{g}\nonumber
\\
L_i&=&0.1\,
\mu\mbox{m}\nonumber\\
r_0&=&0.01\,\mu\mbox{m}\label{eq22}
\end{eqnarray}
where $m$ is the mass of an inclusion,
$r_0$ is the radius of an M-type
inclusion, $L_i$ is the length of an
S-type inclusion, $\kappa$ is the bending rigidity and $\eta$ is the coefficient of viscosity. The range of values
for the surface tension was chosen to be in line with the
discussions presented previously \cite{Evans, Bar}.
\begin{figure}
\begin{center} \epsfig{figure=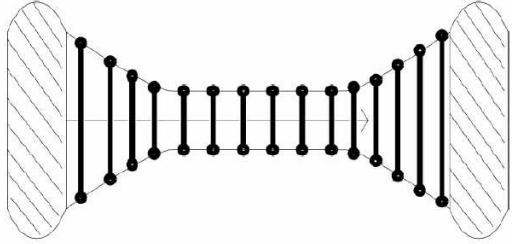,width=7cm}
\end{center}
\caption{\footnotesize Two rod-like embedded (M- type) inclusions
vertically inserted in a fluid biomembrane showing exponentially
decaying thickness-matching constraints on the bilayer at the
boundaries of the inclusions. Heavy solid lines represent
amphiphilic molecules. Figure based on \cite{Dan2}. } \label{fig1}
\end{figure}
\begin{figure}
\begin{center} \epsfig{figure=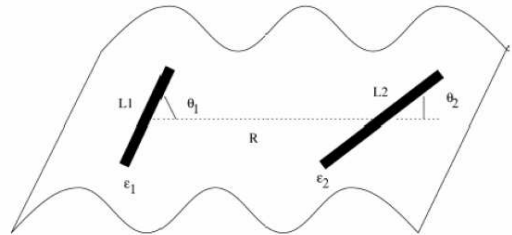,width=7cm}
\end{center}
\caption{\footnotesize Two rod-like surface (S-type) inclusions
placed on the surface of the biomembrane. The rods have lengths
$L_1$ and $L_2$, widths $\epsilon_1$ and $\epsilon_2$ and making
angles $\theta_1$ and $\theta_2$ with the line joining their
centres of mass. Figure based on \cite{Gole}. } \label{fig1}
\end{figure}

To implement the equations of motion, (\ref{eq19}) and
(\ref{eq20}), we first required the computation of $D_t$. The
computation of $D_{\bot}$ part of $D_t$ was based on (\ref{eq12})
for a correlation time of $t=10^{-4}$s. Table I shows the
variation of the $D_{\bot}$ for both the M-type and S-type
inclusions for different values of the surface tension,
$\sigma_0$. It can clearly be seen that the smaller is the surface
tension, the larger is the value of the lateral diffusion
constant.
\begin{center}
\begin{tabular}{|c|c|c|}
\hline
T(K)&$\sigma_0$(J/$\mu$ m$^2$)&D$^{\tiny\mbox{M(S)}}_{\bot}$(m$^2$s$^{-1}$)\\
\hline
300&5$\times 10^{-23}$&0.25$\times
10^{-8}$\\
300&5$\times 10^{-21}$&0.71$\times
10^{-9}$\\
300&5$\times 10^{-18}$&0.46$\times
10^{-10}$\\
\hline
\end{tabular}
\end{center}
\begin{center}
{\footnotesize Table I: Computed values of the lateral diffusion
constant for different values of surface
tension.}
\end{center}

To compute $D_{\|}$ on the basis of (\ref{eq12b}), we first performed a constant-temperature molecular dynamic (MD)-like calculation \cite{Allen} in which the inclusions interacted with each other via the potentials given in (\ref{eq17}) and (\ref{eq18}). These potentials describe the M-M and S-S type interactions  respectively. For the M-S interactions, we adopted the simple mixing rule of an arithmetic average of the M-M and S-S potentials. The temperature was kept constant at $T=300$ K  by applying the Nos\'e-Hoover thermostat (heat bath) \cite{Hashem1}, and the simulation time-step was set at $dt=10^{-9}$s. During this simulation our aim was to obtain the positions of the inclusions as a function of simulation time, and these were recorded at specified time intervals, and the data were then used in (\ref{eq12b}) in a separate simulation code. The same correlation time, $t=10^{-4}$ s, that was used for the computation of $D_{\bot}$ was used for the computation of $D_{\|}$. In this case, this correlation time corresponded to the positions of inclusions recorded after $10^5$ time steps. The value of $D_{\|}$ for the M- and S-type inclusions corresponding to this delay time was obtained to be
\begin{equation}
D^{\tiny\mbox{M(S)}}_{\|}=0.99 \times 10^{-11}\,\,\mbox{m$^2$ s$^{-1}$}.\label{eq22a}
\end{equation}
Comparison of (\ref{eq22a}) and the values listed in Table I shows that the main contribution to the $D_{t}$ comes from the lateral component, which was also found to be comparable to the corresponding value of $D\approx
4.4\times 10^{-9}$ m$^2$s$^{-1}$ obtained from an MD-based
simulation of a fully hydrated
phospholipid dipalmitoylphosphatidylcholine (DPPC)
bilayer diffusing in the $z$-direction
\cite{Tiel}.
\begin{figure}
\begin{center} \epsfig{figure=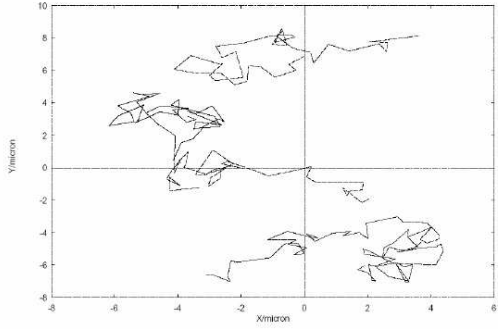,width=7cm}
\end{center}
\caption{\footnotesize Stochastic trajectories of a sample of
S-type inclusions. Both drift and diffusive motions are clearly
visible. } \label{fig1}
\end{figure}
\begin{figure}
\begin{center} \epsfig{figure=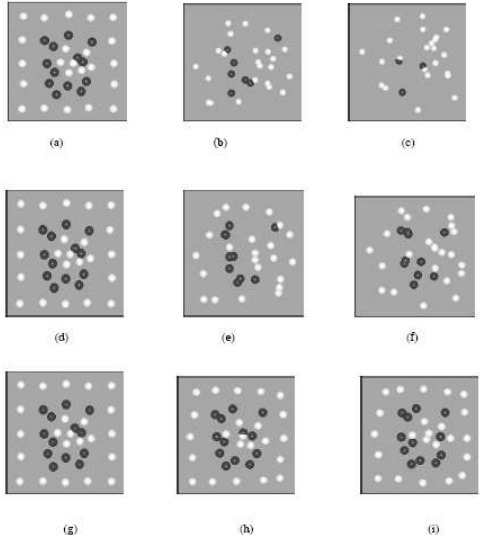,width=7cm}
\end{center}
\caption{\footnotesize A series of snapshots from dynamic
simulations showing the capture of the S-type inclusions (black
spheres) at the cite of the M-type inclusions (white spheres) for
{\it dynamic} M-type inclusions. The top row, (a)-(c), corresponds
to the value of the surface tension $\sigma$=5$\times 10^{-
23}$J/$\mu$m$^2$, the middle row, (d)-(f), to the value $\sigma$=5
$\times 10^{-21}$J/$\mu$m$^2$, and the bottom row, (g)-(i), to the
value $\sigma$=5$\times 10^{- 18}$J/$\mu$m$^2$. Only the centres
of mass of the inclusions are shown.} \label{fig1}
\end{figure}
The choice of correlation
time, i.e. $t=10^{-4}$s, over which the values
of the components of $D_t$ were computed has to be justified.
We justify this choice by noticing that
the diffusion time scale, $t_D$, for a
stochastic particle of mass $m$ is given
by \cite{Dhont}
\begin{eqnarray}
t_D=\frac{m D}{k_B T}.\label{eq23}
\end{eqnarray}
Since the relevant time scale for Brownian particles is at least, $t_D=
10^{-9}$ s, then in order for the
criterion of long-time dynamics, as embodied
in (\ref{eq15}), to be applicable to our
model, the correlation time, $t$, in
(\ref{eq12}) and (\ref{eq12b}) over which the diffusion
constants were calculated, must satisfy the
condition \cite{Dhont}
\begin{eqnarray}
t\gg t_D.\label{eq24}
\end{eqnarray}
For the data given in (\ref{eq22}) and
the values of $D_t$ as computed from any of the relationships in given (\ref{eq14a}), the
values of $t_D$ obtained from (\ref{eq23}) turn out to be much smaller than $t$ indicating
setting the value of the correlation time at
$t=10^{-4}$ was quite reasonable.

A set of dynamic simulations, based on (\ref{eq19}) and (\ref{eq20}), were
performed for the different combinations given in (\ref{eq14a}) to compute the stochastic
trajectories for both the S-type and M-type inclusions, and the rate of capture of the S-type inclusions at the sites of the M-type inclusions. The
temperature was set at $T$=300 K . The
simulation time step, $dt$, in
(\ref{eq19}) and (\ref{eq20}), was set at
$dt=10^{-9}$ s, and each simulation was
performed for a
total time of 4000 $\mu$s, i.e. $4\times
10^{6}$ iterations. The total number of
inclusions consisted of 13 S-type and 23
M-type. Table II shows the variation of the number of S-type inclusions captured at the site of M-type inclusions for different values of the surface tension and different combinations of $D_t$, with the values of $D_{\bot}$ and $D_{\|}$ given in Table I and (\ref{eq22a}). The data in this table refer to the capture of the S-type inclusions when both the S-type and the M-type inclusions were mobile.
\begin{center}
\begin{tabular}{|c|c|c|c|}
\hline
Simulation data&$D_{t}=\frac{1}{2}(D_{\bot}+D_{\|})$&$D_{t}=\frac{1}{3}(2D_{\bot}+D_{\|})$&$D_{t}=\frac{1}{3}(D_{\bot}+2D_{\|})$ \\
&captured S-type&captured S-type&captured S-type\\
\hline
$\sigma_0=5\times 10^{-23},\,\, T=300$&10&11&8 \\
$\sigma_0=5\times 10^{-21},\,\, T=300$&3&4&3\\
$\sigma_0=5\times 10^{-18},\,\, T=300$&1&1&1 \\
\hline
\end{tabular}
\end{center}
\begin{center}
{\footnotesize TableII: Variations of the number of captured S-type inclusions with surface tension and temperature for different  different combinations of the lateral ($D_{\bot}$) and longitudinal ($D_{\|}$) diffusion coefficients. $\sigma_0$ is in units of J/$\mu$m$^2$, and the $D$s are in units of m$^2$ s$^{-1}$. Both the S-type and the M-type inclusions were mobile.}
\end{center}
To make a comparison with the case  when the M-type inclusions were static and only the S-type inclusions were mobile,  the variation of the number of captured S-type inclusions for different combinations of $D_t$ was computed, and the results for a particular value of $\sigma_0$ are listed in Table III.
\begin{center}
\begin{tabular}{|c|c|c|c|}
\hline
Simulation data&$D_{t}=\frac{1}{2}(D_{\bot}+D_{\|})$&$D_{t}=\frac{1}{3}(2D_{\bot}+D_{\|})$&$D_{t}=\frac{1}{3}(D_{\bot}+2D_{\|})$ \\
&captured S-type&captured S-type&captured S-type\\
\hline
$\sigma_0=5\times 10^{-21},\,\, T=300$&3&3&2\\
\hline
\end{tabular}
\end{center}
\begin{center}
{\footnotesize Table III: Variations of the number of adsorbed S-type inclusions for different combinations of the lateral ($D_{\bot}$) and longitudinal ($D_{\|}$) diffusion coefficients when only the S-type inclusions were mobile.}
\end{center}
Fig.3 shows the stochastic trajectories of a set of six S-type inclusions obtained from (\ref{eq19}) and (\ref{eq20}). Both drift motion, represented by the second term on the RHS of (\ref{eq19}), and random diffusive, motion coming from membrane fluctuations, superimposed on the drift motion are visible.

Fig.4 shows the snapshots of a
small portion of the biomembrane for a
simulation in which both the M-type
inclusions  and the S-type inclusions
were allowed to diffuse. In this figure, the black
spheres represent the S-type inclusions
and the white spheres the M-type
inclusions. The spheres represent the
centres of mass of the inclusions. The
snapshots (a-c) refer to the
case with $\sigma_0=5\times 10^{-
23}$J/$\mu$m$^2$, (d-f) to the case with $\sigma_0=5\times 10^{-
21}$J/$\mu$m$^2$, and (g-i) to the case with $\sigma_0=5\times 10^{-
18}$J/$\mu$m$^2$. In each row of this
figure, the first
snapshot, e.g (a), refers to the initial
state, the second snapshot, e.g. (b), to the
state after 2000 $\mu$s and the last snapshot, e.g. (c), to the state
after 4000 $\mu$s. In the
initial state, e.g. (a), in this figure, the outer M-type
inclusions were distributed regularly and
the inner ones were distributed at
random. For the S-type inclusions, these
were all distributed completely at
random. As can be clearly seen from the
snapshots in Fig.4, the smaller the value of
the surface tension, i.e. the top row
snapshots, the higher the rate
of capture of the S-type inclusions by the
M-type inclusions. Furthermore, we
observed that for the higher value of the
surface tension, i.e. the bottom row, there was a clear {\it
collective} motion of the S-type
inclusions, i.e. the surface tension
constrained the motion of the individual
inclusions to an organised type of motion.

The bottom-row snapshots
showed that neither type of inclusions
made any significant diffusion during the
simulation time, i.e. for this
value of the surface tension, the stiffness
of the patch practically prevented any
appreciable movement of the inclusions.
This implied that the biomembrane surface
behaved in a more stiff manner than the
case of low surface tension or the absence
of surface tension as reported in
\cite{Hashem}.
The last snapshot in the first row of this
figure shows that the majority of the
S-type inclusions were captured by the
M-type inclusions, and that there was a
total disorder in the pattern of the M-
type inclusions. This is similar to the results in our previous
simulation in the absence of surface
tension \cite{Hashem}, although a
bigger number of inclusions were
captured during that simulation. This
implies that as the surface tension
becomes smaller and smaller, the
biomembrane behaves more like a fluid, hence
allowing an easier diffusion for any type
of inclusions. This can have direct
practical implications, namely, for the
passage of external particles into the
interior of a cell, the magnitude of the
surface tension can play a very crucial
role. A flexible biomembrane allows more
external particles to arrive at the site of
the embedded proteins (i.e. the M-type
inclusions), hence increasing the rate of
transport of material to the interior of
the cell. This conclusion can have useful
implications for the design artificial membranes, smart
drugs and a better drug delivery
mechanism.

To sum up, our computational
simulation has been able to reveal the
essential role played by the surface
tension in the overall dynamics of the
biomembrane. This is compatible with the
experimental finding \cite{Bar} in which
it was found that for a surface tension
value of $\sigma\ge5\times 10^{-
18}$J/$\mu$m$^2$ the vesicle becomes
quite taut inducing an organised type of
motion to the contents of the vesicle.
In our previous simulation
\cite{Hashem} it was shown that the
rate of capture of the S-type inclusions
was clearly affected by an increase in the
temperature of the biomembrane. We can
now add that this rate is also affected
strongly by the presence of the surface
tension in the biomembrane.

Another finding from our simulation is that the contribution of the lateral component of the diffusion coefficient, $D_{\bot}$, computed from the stochastic fluctuations of the height function of the biomembrane, far outweighs that of longitudinal coefficient, $D_{\|}$. In view of the fact that the larger is the value of the surface tension, the smaller the value of the diffusion coefficient becomes, then if the contribution of surface tension had also been included in the computation of the longitudinal component of the diffusion coefficient, this would have further reduced the value of this component and made its relative importance, vis-a-vise $D_{\bot}$, even less important.


\begin{thebibliography}{99}
\bibitem{Hunt}J.H. Hunt, Foundations of
Colloid Science, Clarendon Press, Oxford,
1993.
\bibitem{Seifert1} U. Seifert, Adv. in
Phys. 46(1997)13.
\bibitem{Nico}S.J. Singer, G.L Nicolson,
Science 175 (1972) 720.
\bibitem{Hashem} H. Rafii-Tabar H and
H.R. Sepangi, Comp.Mat. Sci. 15 (1999)
483.
\bibitem{Dan1}N.Dan, P. Pincus, S.A.
Safran, Langmuir 9 (1993) 2768.
\bibitem{Dan2}N. Dan, A. Berman, P.
Pincus, S.A. Safran, J. Phys. II France 4
(1994) 1713.
\bibitem{Mansfield}E.H. Mansfield, H.R.
Sepangi, E.A. Eastwood, Phil. Trans. R.
Soc. Lond. A 355 (1997) 869.
\bibitem{Bloom}M.Bloom, E.Evans,
O.G.Q Mouristen, Rev. Biophys. 24
(1991) 293.
\bibitem{Abney} J.R. Abney, J.C. Owicki,
in : W. De Pont (Ed), Progress in Protein-
Lipid Interactions, Elsevier, New York
1985.
\bibitem{Marcelja} S. Marcelja, Biophys.
Acta 455 (1976)1.
\bibitem{Owicki} J.C. Owicki, H.M.
McConnell, Proc. Nat. Acad. Sci. USA
76(1979)4750.
\bibitem{Fattal}D.R. Fattal. A. Ben-
Shaul, Biophys. J. 65 (1993) 1795.
\bibitem{Huang}H.W. Huang, Biophys.
J. 50 (1986) 1061.
\bibitem{Stat} D. Nelson, T. Piran and S.
Weinberg, Statistical Mechanics of
biomembranes and Surfaces, Jerusalem
Winter School for Theoretical Physics,
Vol 5, World Scientific, Singapore, 1989.
\bibitem{Safran1} S.A. Safran,
Statistical Thermodynamics of Surfaces,
Interfaces and Biomembranes, Addison-
Wesley, Reading, MA, 1994.
\bibitem{Gompper}G.Gompper, M.
Schick, Self-assembling Amphiphilic
Systems, Phase Transition and Critical
Phenomena, in : C. Domb, J. Lebowitz
(Eds), Academic Press, London, 1994.
\bibitem{Seifert2} U. Seifert, Z. Phys. B
97 (1995) 299.
\bibitem{Alberts} B. Alberts {\it et al}.,
Molecular Biology of the Cell,
Garland, New York, 1994.
\bibitem{Sens} P. Sens, S.A. Safran,
Europhys Lett. 43 (1998) 95.
\bibitem{Evans}E. Evans, W. Rawicz,
Phys. Rev. Lett. 64 (1990) 2094.
\bibitem{Bar}R. Bar-Ziv, T. Frisch, E.
Moses, Phys. Rev. Lett. 75 (1995) 3481;
D. Moroz, P. Nelson, R. Bar-Ziv, E.
Moses,
Phys. Rev. Lett. 78 (1997) 386.
\bibitem{Can}P.B. Canham, J. Theor.
Biol. 26 (1970) 61.
\bibitem{Hel} W. Helfrich, Z. Naturforsh,
28c (1973) 693.
\bibitem{Mathematica} The Mathematica Version 5 (2003), Wolfram Research, Inc. Champaign Ill USA.
\bibitem{Chiu}S.W. Chiu, M. Clark, V.
Balaji, S. Subramaniam, H.L, Scott, E.
Jakobsson, Biophys. J. 69 (1995) 1230.
\bibitem{Haile} J.M. Haile, Molecular Dynamics Simulation: Elementary Methods, Wiley, N.Y., 1992.
\bibitem{Dhont}J.K.G. Dhont, An
Introduction to Dynamics of Colloids,
Elsevier, Amsterdam, 1996.
\bibitem{Ito}C.W. Gardiner, Handbook of
Stochastic Methods,
Springer, Berlin, 1990.
\bibitem{Allen}M.P Allen, D.J. Tildesley,
Computer Simulation
of Liquids, Clarendon Press, Oxford,
1987.
\bibitem{Goul1}M. Goulian, R. Bruinsma
Pincus P., Europhys.Lett. 22 (1993) 145.
\bibitem{Goul2}M. Goulian, R. Bruinsma
Pincus P., Europhys.Lett. 22 (1993)
155E.
\bibitem{Gole} R. Golestanian,M.
Goulian, M.  Kardar M., Europhys. Lett.
33 (1996) 241.
\bibitem{Hashem1}H. Rafii-Tabar , Phys.
Rep. 325 (2000) 239.
\bibitem{Tiel}D.P. Tieleman, H.J.C
Berensden, J. Chem. Phys. 105 (1996)
4871.
\end{thebibliography}
\end{document}